# Pressure-Induced Martensitic Phase Transformation and Microstructure Evolution in nanograined Fe-7%Mn Alloy

Mrinmay Sahu[1*], Sorb Yesudhas[1], Valery I. Levitas[1,2*], and Dean Smith[3]

[1]Department of Aerospace Engineering, Iowa State University, Ames, Iowa 50011, USA

[2]Department of Mechanical Engineering, Iowa State University, Ames, Iowa 50011, USA

[3]HPCAT, X-ray Science Division, Argonne National Laboratory, Argonne, Illinois 60439, USA

*Corresponding authors. Email: msahu@iastate.edu, vlevitas@iastate.edu

## Abstract

The Fe-Mn-based alloys are receiving immense attention due to their applications in the third generation of advanced high-strength steels, owing to their high strength and ductility. A detailed in situ high-pressure structural phase transformation and microstructural evolution in nanograined Fe-7%Mn alloy has been performed using the axial synchrotron X-ray diffraction technique. The ambient BCC phase of Fe-7%Mn undergoes pressure-driven structural PT to the HCP phase at 11.4 GPa. Both BCC and HCP phases coexist up to 15.9 GPa; thereafter, they transform into a pure HCP phase, which remains stable up to the maximum pressure of 30.3 GPa. The XRD study reveals that the $(110)_b$ dense crystallographic plane of the BCC lattice transforms into a densely packed $(002)_h$ peak of the HCP lattice following the orientational relationship $(110)_b \parallel (0001)_h$ via diffusionless Burger's martensitic crystallographic PT pathway. The evolution of crystallite size and microstrain with pressure shows a distinct change during the structural PT. The microstrain exhibits a sharp anomaly at around 10 GPa, suggesting that the microstructural changes precede the structural PT.

## Introduction

Iron (Fe) is one of the most abundant elements on Earth, playing a central role in modern technology and being the principal component of the Earth's core. The phase transformations (PTs) in iron and iron-based alloys driven by pressure and temperature have been extensively studied [1-11]. The pressure-temperature phase diagram of iron can be significantly altered by alloying, leading to changes in structural stability, phase transformation behavior, magnetic and electronic properties, etc. [12, 13]. Among ferroalloys, manganese (Mn) is a particularly important alloying element owing to its high solid solubility in iron, which arises from the similar valence electronic configurations of Mn ($3d^5 4s^2$) and Fe ($3d^6 4s^2$) [14, 15]. Alloying Mn with iron significantly influences magnetic ordering, phase stability, and the temperatures and pressures of phase transitions. As a result, a rich pressure, temperature, and composition-dependent phase transition behavior of Fe alloys can be achieved by tuning their microstructure. In particular, the stability of close-packed phases in Fe is known to increase with the addition of Mn [12-16]. Alloying steels with medium Mn has important applications in the third generation of advanced high-strength steels due to their high strength and ductility [17]. The Mn-alloyed steel alters its microstructural properties, increasing its strength, ductility, and toughness [18].

At ambient conditions, pure $α$-Fe crystallizes in a ferromagnetic body-centered-cubic (BCC) structure. Upon compression, it undergoes a diffusionless reconstructive type PT to the hexagonal close-packed (HCP, $ε$-Fe) structure at around 13 GPa, accompanied by the loss of magnetic order [1, 19-27]. This BCC-to-HCP reconstructive martensitic type phase transformation is generally described by the Burgers mechanism, which involves a coupled lattice shear and atomic shuffle of adjacent planes [28]. Similar pressure-induced BCC-to-HCP transformations have been reported in Fe-Mn alloys, with the transition pressure and coexistence range strongly dependent on Mn concentration [12-16, 29]. Notably, it has also been reported that the high-pressure close-packed phase in Fe-Mn alloys can be retained at ambient pressure, highlighting the strong influence of Mn on phase stability [13]. A few studies on the structural phase transformation of Fe-Mn under shock compression have been reported [13, 30]. To the best of our knowledge, no systematic, detailed high-pressure structural and microstructural studies on Fe-Mn alloys using synchrotron XRD have been reported.

In this work, we investigate the detailed structural and microstructural behavior of nanograined Fe-7%Mn alloy under hydrostatic compression at room temperature using in situ high-pressure synchrotron X-ray diffraction. The equations of state for both BCC and HCP phases are obtained using the second Birch-Murnaghan equation of state (EOS). The phase fraction, crystallite size, and microstrain (MS) are studied in detail.

**Experimental details**

The Fe-7%Mn alloy, with initial dimensions of 36.8 mm × 25.7 mm × 14.9 mm, was prepared at the Army Research Laboratory [30]. The sample thickness was first reduced from 14.9 mm to ~5 mm by cutting the sample into three pieces using electrical discharge machining (EDM) at the Ames National Laboratory machine shop. The thickness was further reduced to 280 µm by cross-rolling on a 100 T rolling mill at the same facility, to obtain a nanograined structure. A small speck, approximately 40 µm in diameter, was removed using a diamond file for a high-pressure (HP) X-ray diffraction study. HP experiments were conducted using a symmetric diamond anvil cell (DAC) equipped with diamond anvils having a 400-µm culet diameter. A rhenium gasket was preindented to a thickness of 61 µm, and a 220 µm hole diameter was drilled at its center using the laser drilling facility at HPCAT. The gasket with the central hole was placed between two opposing diamond anvils to form the sample chamber.

A small piece of nanograined sample, together with tiny ruby chips (approximately 3-5 µm in size), was loaded into the sample chamber. The ruby particles and Helium (He) were used as a pressure sensor and pressure-transmitting medium (PTM), respectively. The sample was compressed inside the DAC using a membrane-driven system at HPCAT. The high-pressure X-ray diffraction (XRD) experiments were performed at the 16-ID-B beam line of HPCAT at the Advanced Photon Source using an X-ray wavelength of 0.4246 Å. The incident X-ray beam was focused to a spot size of 1.3 µm × 1.9 µm onto the sample. The diffracted X-rays were collected using a PILATUS3 X 2M CdTe detector positioned normal to the beam. The sample-to-detector distance was calibrated using the standard XRD pattern of $CeO_2$. Acquired two-dimensional (2D) diffraction images were converted to intensity vs 2-theta using DIOPTAS software [31]. Rietveld refinements of the XRD patterns were performed using the Materials Analysis Using Diffraction (MAUD) [32] program to get the unit cell parameters, volume fractions, and microstructures.

**Results and discussion**

The unit cell of BCC (SG: $Im\bar{3}m$) and HCP (SG: $P6_3/mmc$) pure Fe lattices are shown in Fig. 1. The Wyckoff positions of BCC and HCP phases are 2a (0, 0, 0) and 2c (1/3, 2/3, ¼). We incorporated 7% Mn in the Fe site to match the peak intensities in the Rietveld refinement. Figs. 2a and b exhibit high-pressure XRD patterns of Fe-7%Mn alloy up to 30.3 GPa for selected data points under hydrostatic compression. The typical XRD pattern at 0.4 GPa for the BCC phase show four distinct Bragg peaks $(110)_b$, $(200)_b$, $(211)_b$ and $(220)_b$ (Fig. 2a). The XRD peaks at 0.4 GPa show significant broadening due to the formation of dislocations, grain refinement, and twinning during the pre-deformation caused by the cross rolling and pre-indentation of the sample using diamond anvils. The calculated lattice parameter of the BCC phase, obtained through Rietveld refinement at 0.4 GPa, is 2.8702(9) Å. As pressure is increased, the peaks are shifted towards higher 2-theta angles due to a decrease in *d*-spacing. The ambient BCC phase of Fe-7%Mn is stable up to ~10.5 GPa. Upon further compression, we observed two shoulder peaks $(100)_h$ and $(101)_h$ on either side of high intense $(110)_b$ peak, which is a clear indication of BCC→HCP structural PT at ~11.4 GPa. With even a small compression by 0.7 GPa ($p$ = 12.1 GPa), the intensities of these peaks are significantly enhanced by a factor of 6. In addition, we observed five more distinct peaks $(102)_h$, $(110)_h$, $(103)_h$, $(112)_h$, and $(201)_h$ of the HCP phase at 12.1 GPa. The XRD pattern in Fig. 2a reveals a mixed phase. The Rietveld refinement reveals that $(110)_b$ of the BCC phase is a mixture of $(110)_b$ and $(002)_h$ peaks. At ~16.4 GPa, all the peaks of the BCC phase disappear, and the $(110)_b$ peak of the BCC phase is completely transformed into the $(002)_h$ peak of the HCP phase. The obtained lattice parameter of the pure HCP phase is $a$ = 2.4630(3), $c$ = 3.9591(4) at 16.4 GPa. The intensity of the $(002)_h$ peak of the pure HCP phase at 15.5 GPa and above did not fit well due to the preferred orientation effect. The observed texture of the $(002)_h$ reflection indicates a strong *c*-axis alignment of HCP Fe-7%Mn grains, which is fitted well with the March-Dollase preferred orientation model. The high-pressure HCP phase is stable up to the maximum pressure of 30.3 GPa achieved in this study. After decompression, a reversible HCP to BCC PT is obtained, similar to the parent Fe [5].

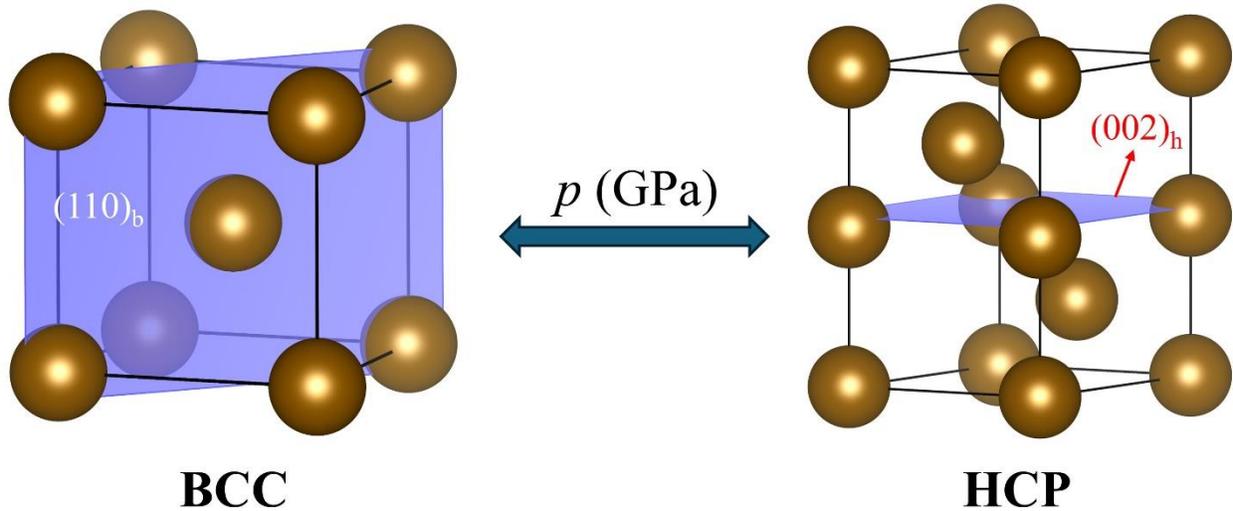

**Fig. 1.** Representation of the crystallographic BCC to HCP phase transformation pathway in Fe-7%Mn. To correlate the BCC (Wyckoff Positions: 2a (0, 0, 0)) unit cell with the HCP, we used the conventional HCP structure with Wyckoff positions, 2a (0, 0, 0) and 2c (1/3, 2/3, 1/4).

A detailed XRD analysis has been carried out to understand the deformation pathway for the BCC to HCP PT in Fe-7%Mn alloy in comparison with Fe. From Fig. 2, the transformation of $(110)_b$ dense crystallographic plane of BCC lattice into a densely packed $(002)_h$ peak of HCP lattice confirms the $(110)_b \parallel (0001)_h$ and $<111>_b \parallel <11\bar{2}0>_h$ orientation relationship between the BCC and HCP phases in accordance with Burger's crystallographic PT pathway. The crystallographic correspondence underlying this transformation pathway is visualized by the $(110)_b$ and $(002)_h$ planes, as depicted in Fig. 1. The BCC to HCP PT involves 12 Burgers variants of $\{110\}_b$, of these, there are six crystallographically equivalent planes of ((110), ($\bar{1}$10), ($\bar{1}$01), (101), (011), and (0$\bar{1}$1)), and each {110} plane has two <111>$_b$ directions. The BCC to HCP PT generates 12 orientation variants of HCP relative to the parent 12 $\{110\}_b$ planes [33]. As a result, the PT is accompanied by the appearance of a preferred orientation in the $(002)_h$ peak produced by the transformation strain and HCP variants, and the $(002)_h$ peak fitted well with the March-Dollase texture model (Fig. 3). The BCC lattice transformed into the HCP lattice through shear of $\{110\}_b$ planes, followed by shuffles of atoms along the <111>$_b$ direction. The shear of $\{110\}_b$ planes induces stretching and compression of interplanar spacings; as a result, the $(102)_h$ plane is stretched with respect to the $(200)_b$, and the $(103)_h$ plane is compressed relative to the $(211)_b$. Consequently, the dense $(110)_b$ planes with AAA stacking transform into the ABAB packing of $(002)_h$ HCP basal planes as seen in the XRD pattern via a diffusionless and systematic reconstructive type Burger's martensitic type PT pathway. Further experiments and theoretical calculations are needed to elucidate the mechanism of the aforementioned phase transformation.

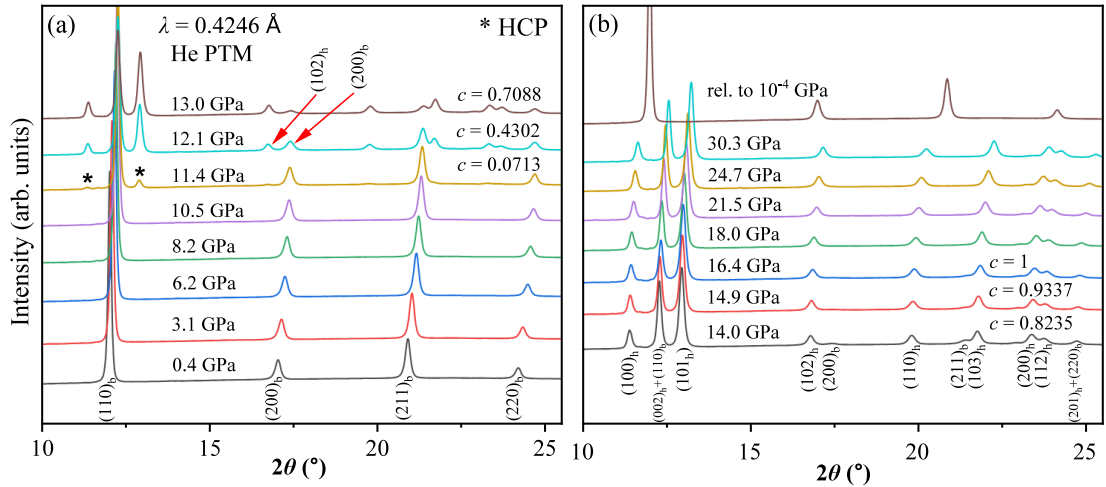

**Fig. 2.** X-ray diffraction patterns of Fe-7%Mn alloy at selected pressures. $c$ represents the volume fraction of the HCP phase.

The evolution of volume fraction of the HCP phase ($c$) with pressure has been studied in detail using Rietveld refinement. The calculated $c$ value is 0.0713 (7.13 %) at 11.4 GPa. The volume fraction $c$ significantly increased from 0.0713 at 11.4 GPa to 0.4302 at 12.1 GPa. At this pressure, the intensity ratio of the $(102)_h$ and $(200)_b$ doublets is ~0.9258, indicating that the HCP phase fraction is ~7.42% lower than that of the BCC phase relative to an ideal 50:50 ratio, in excellent agreement with the calculated value $c = 0.4302$. At 13 GPa, the material is predominantly HCP with $c = 0.7088$, which is evidenced by the huge enhancement of the intensity of $(102)_h$ peak relative to the $(200)_b$ peak. Upon further increasing pressure, approximately 93% of the BCC ($c = 0.9337$) transforms to the HCP phase at 14.9 GPa. The volume fraction $c$ reaches one at 16.4 GPa, indicating the completion of phase transformation (Fig. 2). The extrapolation of volume fraction vs pressure data suggests that the completion of BCC to HCP PT occur at ~15.9 GPa.

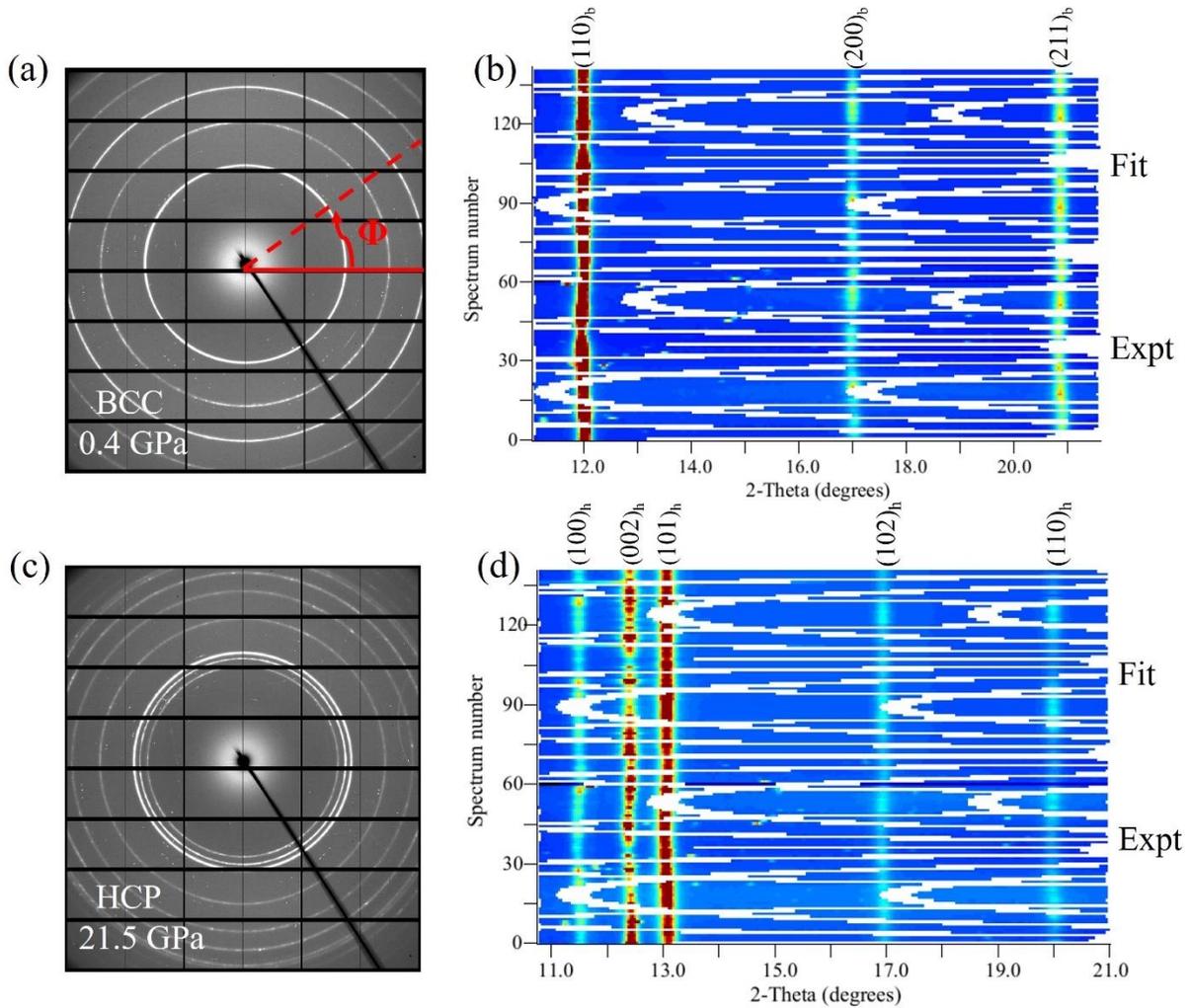

**Fig. 3. Full profile refinements of BCC and HCP phases. (a, c)** 2D XRD images of Fe-7%Mn for BCC (at 0.4 GPa) and HCP (at 21.5 GPa). **(b, d)** The spectra 1-72 are the experimental raw data, and the spectra 73-144 are the full profile refinements of the 2D diffraction images in (a) and (c). The white regions in (b) and (d) are due to gaps in the Pilatus detector. The 2D XRD images were sliced into 72 spectra with a 5° step size.

The pressure dependence of lattice parameters for the ambient and high-pressure phases is shown in Figs. 4a. The $a$-axis of the BCC lattice monotonously decreases with pressure up to 11.4 GPa. Afterwards, the lattice parameters slightly scattered up to 15 GPa, which is due to the coexistence of both the BCC and HCP phases. The $a$ and $c$-axes lattice parameters of the HCP phase monotonously decrease in both the mixed and single phases up to 30.3 GPa. The c/a ratio versus pressure of the HCP phase is shown as the inset in Fig. 4a. The pressure dependence of c/a ratio shows a distinct slope of $-5.2 \times 10^{-4}$ GPa$^{-1}$ up to 15.9 GPa, and it reduces to $-1.4 \times 10^{-4}$ GPa$^{-1}$. A significant slope change in the $c/a$ ratio during the initial growth of the HCP phase is due to the constrained growth of the minority HCP phase within the surrounding BCC phase, resulting in a distortion of the HCP lattice caused by interfacial strain between the two phases. In the mixed-phase region, the axial compressibility along the $c$-axis is higher than that along the $a$-axis, demonstrating anisotropic lattice compression. The equation of state (EOS) for the BCC and HCP phases is obtained by fitting the experimental $p$-$V$ data to the second-order Birch-Murnaghan (BM) EOS. We did not include the unit cell volumes of the mixed phase region in our EOS. The calculated bulk modulus ($B_0$) and its first derivative ($B_0^{'}$) of the BCC and HCP phases are 159.93 GPa and 170.85, respectively. In the HCP phase, a unit cell volume reduction of approximately 4.53% is noticed relative to the BCC phase at the phase boundary, indicating a discontinuous volume collapse associated with the BCC → HCP structural transformation and confirming its first-order nature.

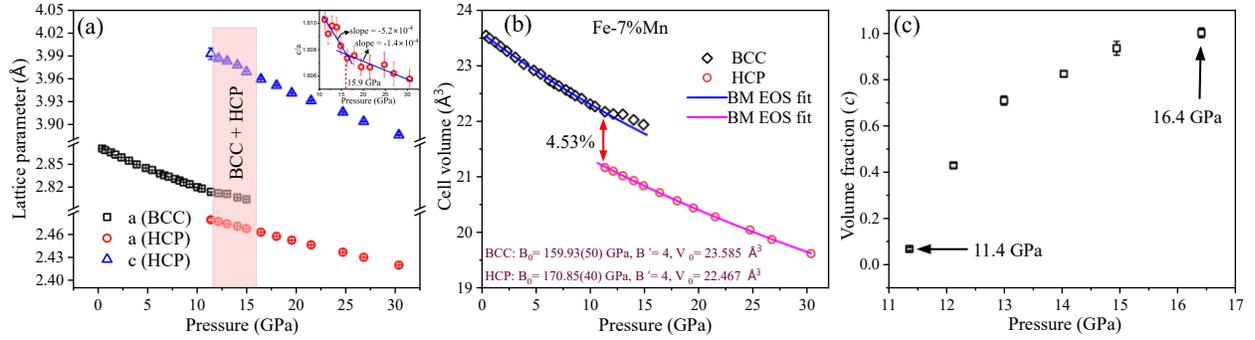

**Fig. 4. Evolution of the unit cell parameters and volume fraction of Fe-7%Mn. (a)** Lattice parameters of BCC and HCP phases. The c/a ratio vs pressure is shown as an inset **(b)** *p-V* EOS of the BCC and HCP phases. Solid lines indicate the second-order Birch-Murnaghan equation of state (BM-EOS) fits to the experimental data. **(c)** The evolution of the volume fraction of the HCP phase.

The pressure dependence of crystallite size (CS) and microstrain (MS) of Fe and Fe-based alloys is not well documented in the literature. Fig. 5a shows the evolution of CS and MS of the BCC phase with pressure. The calculated CS and MS values for the BCC phase at 0.4 GPa are 60.8 nm and 0.0028, respectively. Crystallite size in the BCC phase monotonously decreases with pressure in the whole range, with two plateaus around 5.4 GPa and before PT at ~10.5 GPa. In contrast, the MS increases with pressure, exhibiting a sharp change in the pressure range of 3.2 GPa, and then starts decreasing from 5.4 GPa up to ~10 GPa. The distinct changes discerned in the CS and MS at 5.4 GPa have never been reported in Fe-based alloys before. The origin of this microstructural anomaly, which occurs without a structural phase transition, is unknown. This could be due to the generation of dislocations at the grain boundaries, a change in dislocation core structure resulting in a change in dislocation patterns, reorientation of magnetic ordering caused by the presence of Mn, grain boundary motion, and other factors. Further microstructural studies are required to understand the plausible origin of this phenomenon, which is beyond the scope of this manuscript. The calculated CS of the BCC phase at 10.5 GPa is 45.5 nm, and the MS change observed at 10 GPa is 0.00236. With further compression, the crystallite size increased slightly and then exhibited a sharp decrease. In contrast, MS started increasing from 10 GPa up to 11.4 GPa. In the mixed phase region, the MS dramatically increases from 0.00250 (11.4 GPa) to 0.00358 (15 GPa). Although the initiation of the PT is at 11.4 GPa, distinct CS and MS anomalies are seen at around 10 GPa. This is not surprising, because the microstructural changes originated slightly earlier than the structural PT, which acts as a precursor for the phase transformation seen in the XRD pattern at 11.4 GPa. This observation is in agreement with the X-ray magnetic circular dichroism (XMCD) studies under pressure on Fe, which showed that the magnetic transition appears before the structural PT [34]. The CS and MS changes for the HCP phase with pressure as shown in Fig. 5b. The observed CS and MS values at 11.4 GPa are 40.82 nm and 0.00329, respectively. The CS initially gradually decreases up to 16.4 GPa (29.67 nm); thereafter, it remains nearly constant up to 30.3 GPa. We did not notice any appreciable change in the MS up to 24.6 GPa.

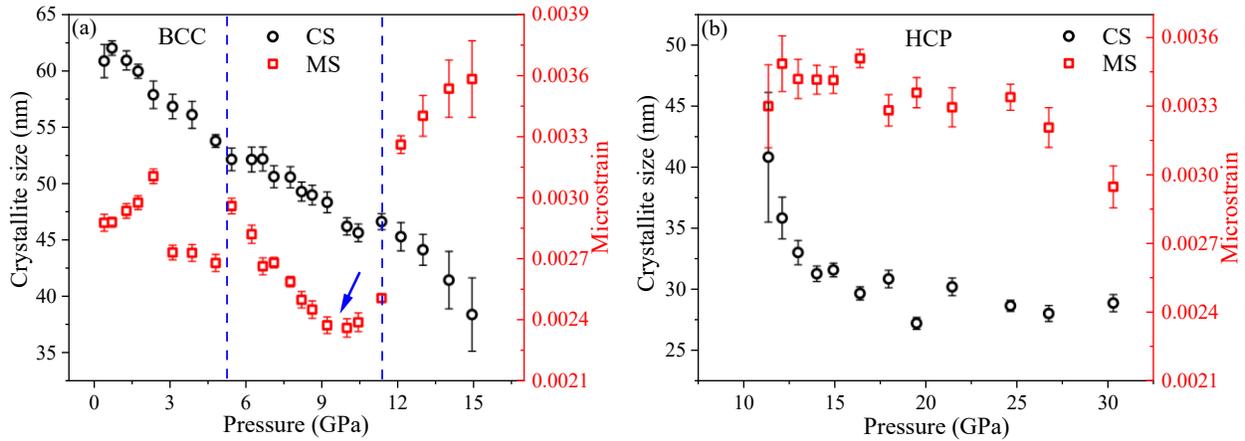

**Fig. 5. Pressure-induced microstructural change in the Fe-7%Mn alloy**. **(a)** Pressure dependence of the crystallite size (CS) and microstrain (MS) for the BCC phase. The vertical blue lines and arrow highlight anomalies observed in MS and CS at 5.4 GPa and just prior to the structural phase transition, respectively. **(b)** Pressure dependence of the crystallite size and microstrain for the high-pressure HCP phase.

In conclusion, we have performed systematic high-pressure synchrotron XRD measurements on Fe-7%Mn alloy up to 30.3 GPa. The BCC to HCP martensitic PT is observed at 11.4 GPa, and the BCC phase completely transforms into the HCP phase at 15.9 GPa through a BCC-HCP mixed phase domain of width 4.5 GPa. A significant decrease in the *c/a* ratio with pressure of the HCP phase in the mixed phase region suggests that a distortion of the HCP lattice is caused by interfacial strain between the two phases. The EOS of BCC and HCP phases is calculated using the second-order Birch-Murnaghan EOS. The CS and MS revealed an unexplored anomaly at 5.4 GPa. The microstructural study indicates that the CS and MS changes of the BCC phase initiated at 10 GPa, much earlier than the structural phase transformation. The microstrain exhibits a discontinuous increase during the BCC-to-HCP phase transition.

**Acknowledgments**

V.I.L. greatly acknowledges ARO (W911NF2420145), NSF (DMR-2246991 and CMMI-2519764), and Iowa State University (Murray Harpole Chair in Engineering). The synchrotron XRD experiments were carried out at HPCAT (Sector 16), Advanced Photon Source (APS), and Argonne National Laboratory. HPCAT operations are supported by DOE-NNSA's Office of Experimental Science. The Advanced Photon Source is a U.S. Department of Energy (DOE) Office of Science User Facility operated for the DOE Office of Science by Argonne National Laboratory under Contract No. DE-AC0206CH11357. We would like to thank Jeffrey T. Lloyd of the DEVCOM Army Research Laboratory for providing the samples.

**Competing interests**

The authors declare no competing interests.

**Author contributions**

M.S. and S.Y. performed the experiments. M.S. post-processed the XRD data. M.S. and Y.S. analyzed the results. V.I.L. conceived the study and supervised the project. M.S., S.Y., and V.I.L. prepared the manuscript.**References:**

[1] T. Takahashi and W. A. Bassett, *High-pressure polymorph of iron*, Science **145**, 483–486 (1964).

[2] D. Andrault, G. Fiquet, M. Kunz, F. Visocekas, and D. Häusermann, *The orthorhombic structure of iron: An in situ study at high temperature and high pressure,* Science **278**, 831 (1997).

[3] R. D. Taylor, M. P. Pasternak, and R. Jeanloz, Hysteresis in the high-pressure transformation of bcc to hcp iron, J. Appl. Phys. **69**, 6126 (1991).

[4] Y. Yoshizawa, S. Oguma, and K. Yamauchi, *New Fe-based soft magnetic alloys composed of ultrafine grain structure,* J. Appl. Phys. **64**, 6044 (1988).

[5] S. Merkel, A. Lincot, and S. Petitgirard, *Microstructural effects and mechanism of bcc–hcp–bcc transformations in polycrystalline iron*, Phys. Rev. B **102**, 104103 (2020).

[6] E. Edmund, D. Antonangeli, F. Decremps, G. Morard, S. Ayrinhac, M. Gauthier, E. Boulard, M. Mezouar, M. Hanfland, and N. Guignot, *The structure and elasticity of cubic Fe–Si alloys at high pressures,* Phys. Rev. B **100**, 134105 (2019).10